\documentclass[twocolumn,aps,prl,longbibliography]{revtex4-1}
\usepackage[english]{babel}
\usepackage{graphicx}
\usepackage{subfigure}
\usepackage{array}
\usepackage{hhline}
\usepackage{amsmath}

\usepackage[dvipsnames]{xcolor}

\newcommand{\RNum}[1]{\uppercase\expandafter{\romannumeral #1\relax}}
\newcommand{\mum}{~$\mu$m{}}

\begin{document}
 
\graphicspath{{./fig_submit/}}

\author{Sarah Benchabane\email{sarah.benchabane@femto-st.fr}, Aymen Jallouli, Laetitia Raguin, Olivier Gaiffe, Jules Chatellier,  Val\'erie Soumann, Jean-Marc Cote, Roland Salut, and Abdelkrim Khelif }
\affiliation{FEMTO-ST, Universit\'e de Bourgogne Franche-Comt\'e, CNRS, UFC\\ 15B Avenue des Montboucons, F-25030 Besan\c con Cedex, France}

\title{Nonlinear coupling of phononic resonators induced by surface acoustic waves}

\begin{abstract}
The rising need for hybrid physical platforms has triggered a renewed interest for the development of agile radio-frequency phononic circuits with complex functionalities. The combination of travelling waves with resonant mechanical elements appears as an appealing means of harnessing elastic vibration. In this work, we demonstrate that this combination can be further enriched by the occurrence of elastic non-linearities induced travelling surface acoustic waves (SAW) interacting with a pair of otherwise linear micron-scale mechanical resonators. Reducing the resonator gap distance and increasing the SAW amplitude results in a frequency softening of the resonator pair response that lies outside the usual picture of geometrical Duffing non-linearities. The dynamics of the SAW excitation scheme allows  further control of the resonator motion, notably leading to circular polarization states. These results paves the way towards versatile high-frequency phononic-MEMS/NEMS circuits fitting both classical and quantum technologies. 

\end{abstract}

\maketitle

The on-chip manipulation of elastic or mechanical vibrations, whether hosted and localized within resonators or traveling in the material substrate, has been leading to significant progress in diverse areas of applied and fundamental science. Surface acoustic wave (SAW) devices are a striking example of such rapid developments. They are ubiquitously used in modern telecommunications systems~\cite{Morgan2007} and are extremely relevant as sensing components in a number of applications~\cite{Go2018,Devkota2017}. 
Micro- and nano-mechanical systems, for their part, have imposed themselves as prevalent means of implementing extremely sensitive sensing devices~\cite{Li_NNano2007,Jensen_NNano2008,Chaste2012,Hanay_NNano2015}. They are characterized by a rich dynamics, involving linear and nonlinear phenomena~\cite{Eichler_NL2011,Villanueva2013,Truitt2013,Mahboob_NL2015} that opens exciting prospects for mechanical signal processing. This dynamics can be further enriched in coupled systems where mechanical mode coupling can be induced in e.g. clamped beams~\cite{Karabalin2008,Okamoto_NPhys2013,Faust_NPhys2013}, pillars~\cite{Raguin_NComm2019,Doster2019} or through intermodal coupling~\cite{Westra_PRL2010,Mahboob_PRL2013,Mathew2016,Matheny_Science2019}. 
Over the past few years, both types of mechanical systems have been attracting renewed interest~\cite{Delsing2019} and were exploited to achieve dynamic and coherent interaction with a variety of physical systems, including holes and electrons~\cite{McNeil2011,Hermelin2011,Hsiao_NComm2020}, quantum dots~\cite{Couto2009,Metcalfe_PRL2010,Kettler_NNano2020,Vogele_AQT2020} spins~\cite{MacQuarrie_PRL2013,Golter_PRX2016,Whiteley_NPhys2019}, superconducting circuits~\cite{OConnell_Nature2010,Gustafsson_Science2014,Satzinger_Nature2018,Bienfait_Science2019} or and photonic devices~\cite{Fuhrmann2011,Balram2016,Patel_PRL2017,Riedinger_Nature2018,Kalaee_NNano2019}. Enhancing control over such mechanical systems at the micro- or nano-scale and in the radio-frequency regime therefore holds promises for the implementation of complex phononics-based hybrid platforms permitting high-speed operation. Recent works have proposed to combine NEMS-based architectures with travelling elastic waves to demonstrate dynamic phononic systems~\cite{Hatanaka_NNano2014}, that could be further enriched by harnessing mechanical nonlinearities~\cite{Cha_NNano2018,Kurosu_NComm2018,Kurosu_PRApp2020}. The reported approaches exploit geometric, electrostatic or wave-mixing induced non-linear behaviors, taking advantage of the high-quality factor resonance of mechanical elements. The current limit of this approach may however lie in the attainable resonant frequencies that are heavily conditioned by NEMS classically operating at maximum frequencies of a few MHz. 

In this work, we demonstrate the possibility to implement a nonlinear phononic platform based on the interaction of travelling SAWs and coupled mechanical resonators capable to operate in the radio-frequency regime.  We analyze a pair of low aspect ratio cylindrical pillars vibrating on a flexural mode and demonstrate the obtaining of unexpected non-linear coupled mechanical states from these otherwise linearly-behaving mechanical resonators. The observed experimental results, supported by analytical modelling and numerical simulations, show that these non-linearities are the result of an interplay between the resonator mode symmetry and the elastic field distribution on the substrate surface and that they exhibit a strong dependence on the resonator gap distance and the SAW excitation scheme. The elastic energy distribution affects in turn the substrate surface displacement, hence disturbing the resonators dynamics and leading to the occurrence of circular state of motion. The proposed devices, that operate in the 70-MHz range, are readily scalable to higher frequencies. They illustrate the relevance of SAW-based phononic devices for the implementation of high-frequency electromechanical circuits with complex dynamics.

\section{Results}

\begin{figure}[tb]
 \centering\includegraphics[width = 88 mm]{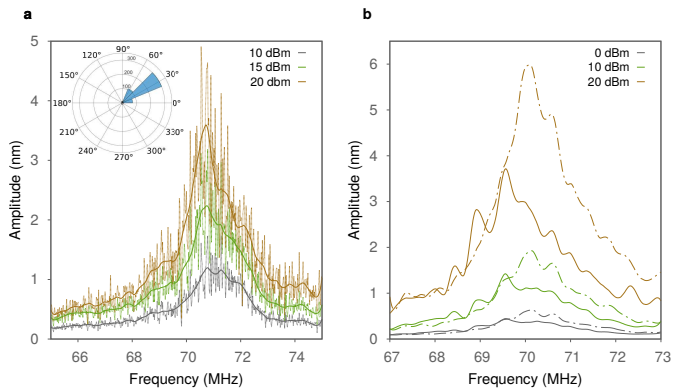}
 \caption{Experimental frequency responses of SAW-driven phononic microresonators. (a) Response of a single pillar with a diameter of 4.4\mum{} and a height of about 4\mum{} excited by a SAW for different input powers applied to the IDT (10~dBm, 15~dBm and 20~dBm). The light dotted lines correspond to data obtained by laser scanning interferometry and acquired every 10~kHz. The solid lines are obtained by applying a Savitsky-Golay smoothing filter to the raw experimental data. Inset: Angular plot of the distribution of the orientation of the flexural mode vibration for frequencies between 65 and 75~MHz. (b) Frequency response for two pillars within a pair with a 6-$\mu$m{} gap distance excited by a longitudinal SAW wave vector under different drive powers. The solid line corresponds to the resonator located closest to the SAW source, the dashed line to the resonator located farthest from the source. 
}\label{fig:singlepillar}
 \end{figure}
 
\subsection{Linear regime: single resonators and large gap distances. }
The phononic resonators under study consist of micron-scale cylindrical pillars with an aspect ratio of the order of one, deposited atop a single-crystal piezoelectric substrate. The resonators are excited by a Rayleigh surface acoustic wave generated by an interdigital transducer harmonically-driven at frequencies about 70~MHz. The resulting SAW amplitude depends linearly on the applied RF power. The mode of interest is a first-order flexural mode, that is theoretically composed of two degenerate, orthogonally polarized eigenstates due to the circular cross-section of the resonators. The frequency response of isolated SAW-coupled resonators were previously shown to exhibit Fano lineshapes, resulting from the interaction of the impinging SAW with the localized mechanical resonator~\cite{Raguin_NComm2019}. Figure~\ref{fig:singlepillar}a reports the out-of-plane vibration amplitude measured out of a single resonator subjected to a SAW driven by radio-frequency (RF) input powers ranging from 10 to 20~dBm. 
A single resonance peak appears at a frequency of 70.7~MHz. The degeneracy lifting expected from fabrication imperfections in resonators with circular symmetries cannot be observed here. This can be accounted for by the low quality factor (\textit{Q}-factor) of the resonance, that is about equal to 30. When increasing the drive power, the vibration amplitude of the phononic resonator shows a predictable linear dependency, as expected from such low quality-factor mechanical structures under the present weak excitation, with a constant resonant frequency and $Q$-factor. 

Similar experiments were conducted on a pillar pair with a 6-$\mu$m gap distance excited by a SAW with a wave vector directed along the inter-resonator axis. Figure~\ref{fig:singlepillar}b shows that increasing the SAW drive power again leaves the frequency position and quality factor of the pillar pair resonances unaffected. The coupling mechanism is therefore here independent on the SAW amplitude, as expected for linearly strain-coupled mechanical resonators. 
\begin{figure*}[t]
 \centering\includegraphics[width =170mm]{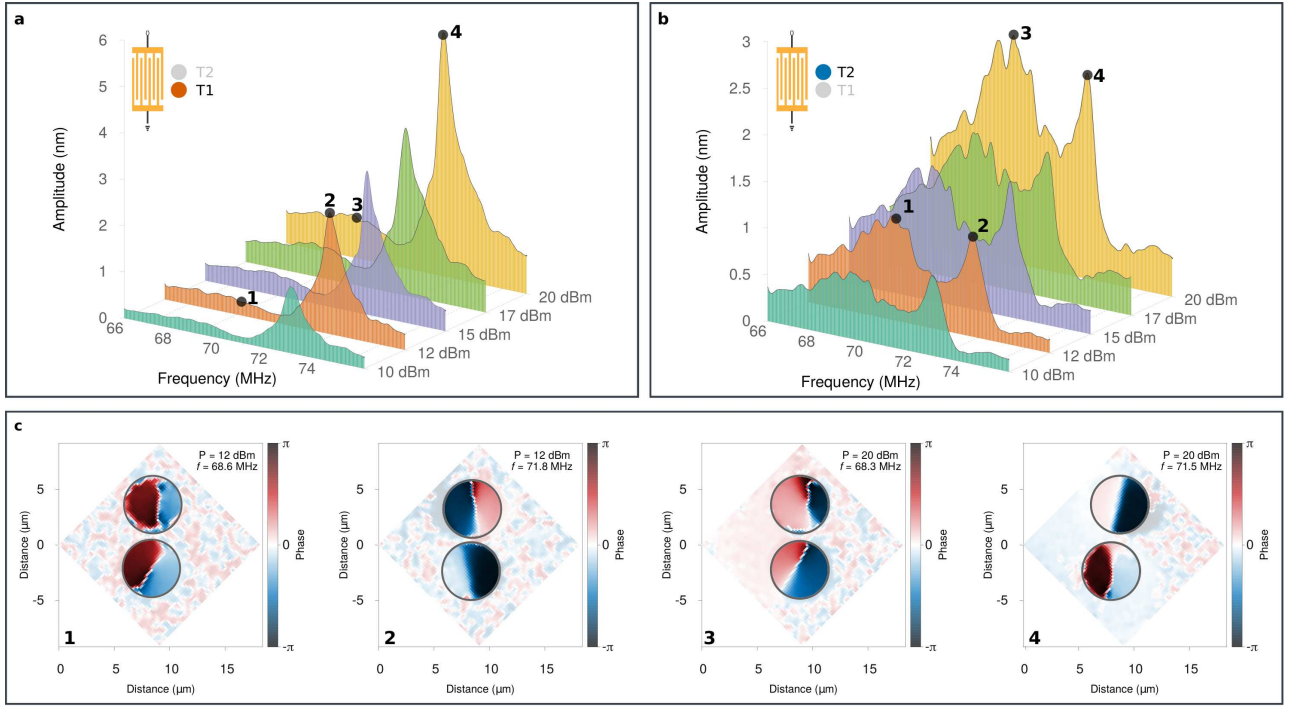}
 \caption{
 Experimental frequency and phase responses of a pillar pair excited by a transverse SAW. The gap distance is 1.5~$\mu$m{}. (a) Response of resonator $T1$. (b) Response of resonator $T2$. (c) Experimental phase maps at 12~dBm for the lower frequency mode (\romannumeral 1) at 68.6~MHz (1) and for the higher-frequency mode (\romannumeral 2) at 71.8~MHz (2);  at 20~dBm for the lower frequency mode (\romannumeral 1) at 68.3~MHz (3) and for the higher-frequency mode (\romannumeral 2) at 71.5~MHz (4).
 } \label{15_transv_nl}
\end{figure*}

\subsection{Frequency softening in smaller gap distance coupled-resonator systems. }
The previous observations do not hold, however, for shorter gap distances, where SAW mediated dipole-like interaction is dominant~\cite{Raguin_NComm2019}. A similar investigation of the resonator response as a function of the drive power was performed for 1.5-$\mu$m-spaced pillar pairs excited by SAWs propagating either along the inter-resonator axis (longitudinal excitation) or in the direction normal to the inter-resonator axis (transverse excitation). 
The results obtained in the case of a transverse excitation are reported in Figure~\ref{15_transv_nl}a and \ref{15_transv_nl}b for each resonator, here denoted $T1$ and $T2$. The resonator pair frequency response reveals the existence of two vibrational modes. The first mode (\romannumeral 1) is found at about 68~MHz and has a quality factor of about 30 that is comparable with the one observed for a single pillar. A higher frequency mode (\romannumeral 2) lying slightly below 72~MHz can also be found, with a quality factor now reaching about 60. The resonance line shape exhibits the lightly asymmetric profile expected in coupled resonator systems. 

Phase maps obtained by laser scanning interferometry and reported in Figure~\ref{15_transv_nl}c reveal that mode (\romannumeral 1) corresponds to a symmetric mode, while mode (\romannumeral 2) corresponds to an antisymmetric mode with respect to the propagation axis. Increasing the drive power in the 10 to 20~dBm range does not significantly affect the overall modal behaviour of the coupled resonator system. It however results in the occurrence of a resonance frequency shift for the higher frequency mode (\romannumeral 2), leading to a downshift of about 400~kHz, while the lower frequency symmetric mode (\romannumeral 1) remains at the same frequency position.

\begin{figure}[!htbp]	
	\centering
		\includegraphics[width=88mm]{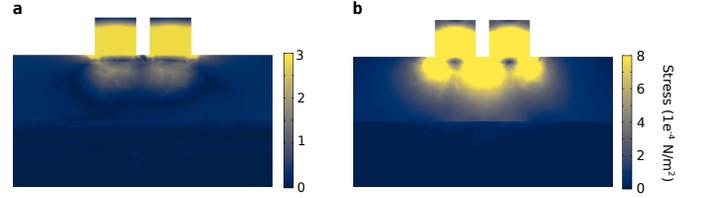}
	\caption{Finite element method simulations of the Von Mises stress distribution. The two 1.5~$\mu$m{}-spaced resonators are excited by a linear elastic line source. (a) Symmetric mode (\romannumeral 1), here found at 70.73~MHz. (b) Anti-symmetric mode (\romannumeral 2) at 73.34~MHz. }\label{VonMises}
\end{figure}

The observed behavior therefore hints at a nonlinear coupling of the first-order flexural modes of the two phononic resonators, despite each individual resonator exhibiting an otherwise linear response. 
The involved non-linear coupling mechanism is therefore different from those usually exploited in NEMS, that rely on the coupling of high-quality factor mechanical resonators with geometrical non linearities~\cite{Nayfeh1995}. The higher quality factor obtained by resonator-to-resonator coupling could help triggering non linear interactions by lowering the amplitude thresholdrequired to reach the critical point. However, these interactions here result in a frequency downshift with increasing drive amplitude, while geometrical nonlinearities are expected to result in a frequency upshift in such clamped-free resonators. 
Such coupling-induced softening nonlinearities were previously reported in phononic crystals made of local resonators coupled by nonlinear graphene membranes~\cite{Midtvedt_NComm2014}. 
In our case, coupling occurs through the substrate surface and is mediated by the SAW elastic energy distribution, that is itself conditioned by the resonator-to-resonator coupling conditions. The observed non-linear behavior can be described as resulting from a softening of the coupling spring constant induced by the highly-confined SAW. This effect may allegedly be understood by considering the coupled pillars as forming a cavity for SAW with a strong energy confinement in the 1.5-$\mu$m-wide gap. The system becomes resonant, hence promoting energy transfer between the two resonators as the SAW amplitude increases. Finite element method (FEM) simulations based on a linear elastic model confirm that the phononic resonators induce a strong localization of the elastic energy distribution at the resonator vicinity. The Von Mises stress distributions reported in Figure~\ref{VonMises} further highlight that stress localization within the resonator gap is only observed for the anti-symmetric mode (\romannumeral 2). This observation is in good in agreement with the experimental device response, as only mode (\romannumeral 2) experiences a nonlinear frequency downshift. 

\begin{figure}[!htbp]	
	\centering
		\includegraphics[width=88mm]{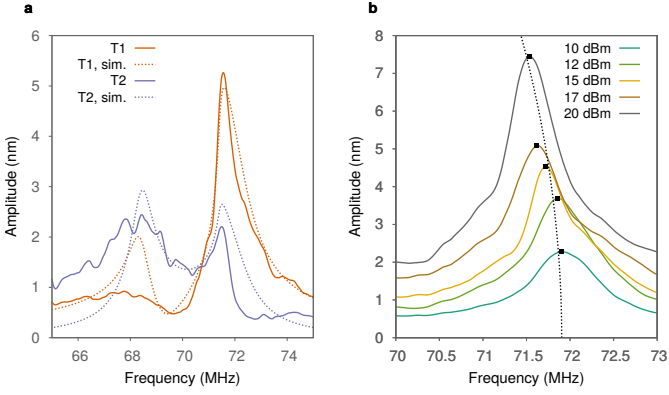}
	\caption{Coupled-oscillator model of a pillar pair excited by a transverse SAW. (a) The solid lines correspond to the experimental data for a pillar pair with a 1.5~$\mu$m{} gap distance excited using a RF input power of 20~dBm. The dashed lines correspond to the theoretical frequency response obtained out of the proposed coupled-oscillator model for an applied force amplitude $F_0=10~\mu$N.  (b) Sum of the measured amplitudes of the two resonators for the anti-symmetric mode. The dots indicate the maximum displacement for each input RF powers, the dashed line is a fitting to the theoretical backbone curve given in equation~\ref{backbone}.}\label{model}
\end{figure}
As a possible description of this SAW-coupled resonator system, we propose to use a tentative simplified model based on two single-mode linear oscillators with respective masses $m_i$ ($i=1,2$) coupled by a non linear coupling strength $k_{nl}$. The corresponding equation of motion can be written as: 
\begin{subequations}
    \begin{align}
        m_1\ddot{x_1} + \gamma \dot{x_1} + k x_1 + \kappa_{12} ( x_1 - x_2) + \kappa_{nl} ( x_1 - x_2 )^3 \nonumber\\ = F_0 \cos(\Omega t),\label{eq_x1}  \\ 
        m_2\ddot{x_2} + \gamma \dot{x_2} + k x_2 + \kappa_{12} ( x_2 - x_1)  + \kappa_{nl} ( x_2 - x_1 )^3 = 0, \label{eq_x2} 
    \end{align}
\end{subequations}
where $x_1$ and $x_2$ are the displacement amplitudes for each resonator, $\gamma$ is the linear damping, $k$  is the resonator spring constant and $\kappa_{12}$ is the linear coupling constant. The unperturbed frequencies of the resonators are defined as  $\omega_{0}^{(i)}=\sqrt{k/m_i}$.
The two effective masses are defined from the physical mass of the resonators ($m_i=\rho V_i$, $\rho$ being the mass density and $V_i$ the volume of the resonators). 
A difference in mass of about 5\%, corresponding to an estimated difference in height of 200~nm between the two resonators, was introduced in the model. The frequency softening effect is introduced by assuming a negative $\kappa_{nl}$ constant. The term on the right-hand side corresponds to the SAW excitation, modelled as an external periodic force of strength $F_0$ and frequency $\Omega$. This external force is applied to only one of the resonators to avoid artificially forcing a relative phase condition between the two oscillators.
Figure~\ref{model}a displays the experimental response of the two resonators for a drive power of 20 dBm, along with the corresponding simulated vibration amplitude. 
The spring constant $k=116$~[kN/m] and the linear coupling constant $\kappa_{12}= 5.3$~[kN/m] used in the model are estimated from the two resonance frequencies measured for mode \textit{(i)} and mode \textit{(ii)} of the pillar pair (see Supplementary Note~\RNum{2}). 
In order to estimate $\kappa_{nl}$, we rewrite the system of equations (\ref{eq_x1}-\ref{eq_x2}) assuming equal resonator masses ($m_1=m_2=m$; Supplementary Note~\RNum{3}):\\
\begin{equation}
    \ddot{u} + \mu \dot{u} + \omega_0^2 u   + \alpha u^3 \\ = F \cos(\Omega t),
    \label{eq_duffing}
\end{equation}
where $\omega_0=\sqrt{\frac{k + 2\kappa_{12}}{M}}$ , $\mu=\gamma/M$ and $\alpha=2\kappa_{nl}/M$. \\$u=x_1-x_2$ is the amplitude difference between the two resonators.
The natural frequency response curve of the nonlinear system can then be described by expressing its backbone curve that follows a typical parabola-shaped curve as a function of the maximum amplitude of $u$:  
\begin{equation} \label{backbone}
    \Omega=\omega_0+\frac{3}{8}\frac{\alpha}{\omega_0}a^2
\end{equation}
where $a$ represents the amplitude of $u$. Fitting the experimental resonance frequency at different drive power to this backbone curve equation, as presented in Figure \ref{model}b, yields a nonlinear coupling constant $\kappa_{nl}$ of the order of -0.016~[kN/m$^3$]. The nonlinear coupling term $\pm \kappa_{nl} ( x_1 - x_2 )^3$ described in equation (\ref{eq_duffing}) as $\pm \kappa_{nl} ( x_1 - x_2 )^3$ affecting only the out-of-phase mode, the proposed model captures the mode symmetry dependence of the observed nonlinear effect, with a symmetrical, in-phase mode frequency position remaining independent of the drive SAW amplitude.

\begin{figure*}[tb]
 \centering\includegraphics[width = 170mm]{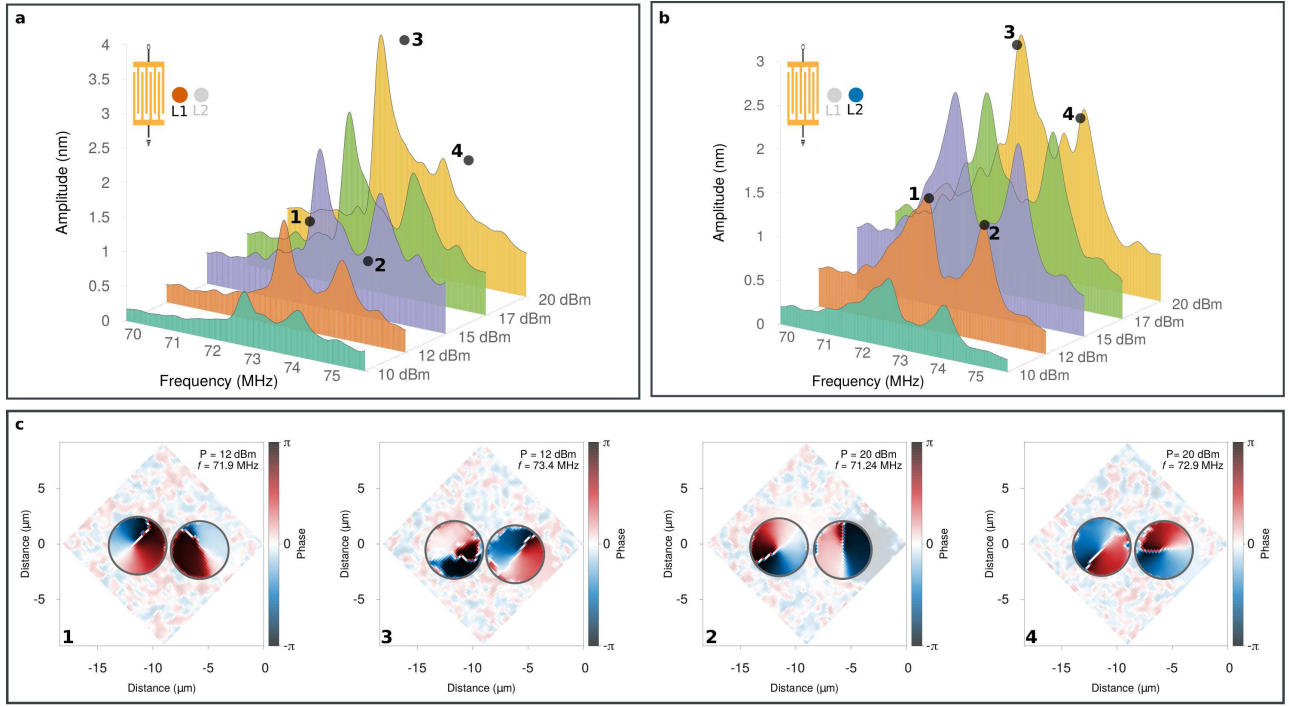}
 \caption{
 Experimental frequency and phase responses of a pillar pair excited by a longitudinal SAW. The gap distance is 1.5~$\mu$m{}. (a) Response of resonator \textit{L1}. (b) Response of resonator \textit{L2}. (c) Experimental phase maps at 12~dBm for the lower frequency mode~(\romannumeral 3) at 71.9~MHz (1) and for the higher-frequency mode (\romannumeral 4) at 73.4~MHz (2); at 20~dBm for the lower frequency mode (\romannumeral 3) at 71.24~MHz (3) and for the higher-frequency mode (\romannumeral 4) at 72.9~MHz (4).
 }
 \label{15_long_nl}
\end{figure*}
\begin{figure}[!htb]
 \centering\includegraphics[width = 88mm]{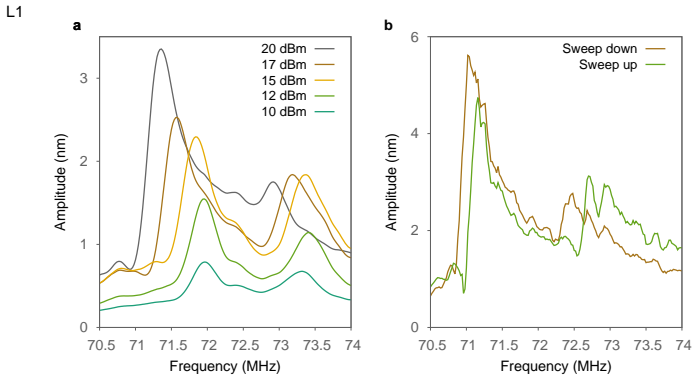}
 \caption{Hysteresis cycle. (a) Frequency response of resonator $L1$ for varying input powers.  (b) Comparison of up and down frequency sweeps for the first resonance of pillar $L1$ (closest to the excitation source) at an RF input power of 20~dBm. The frequency shift linked to the hysteresis cycle appears at about 70.9~MHz.
 } \label{hyst}
\end{figure}

The occurrence of non-linearities was confirmed by investigating a similar pillar pair subjected to a longitudinal excitation. The measured frequency responses, reported in Figure~\ref{15_long_nl}, show that the resonator pair, composed by two pillars $L1$ and $L2$, again hosts two sets of modes appearing at the respective frequencies of 71.5~MHz (mode~((\romannumeral 3) and 73~MHz (mode~(\romannumeral 4)) at 10~dBm. Increasing the applied drive power now leads to a frequency downshift for both resonances. The higher frequency resonance shift is comparable to the one observed in the transverse case and is estimated to be slightly lower than 400~kHz. The resonance frequency of mode~(\romannumeral 4) is also downshifted, with a shift now of the order of 600~kHz, as displayed in Figure~\ref{hyst}a. But one of the most interesting feature of this frequency response lies in the observed change in resonance profile, that presents an increasing asymmetry as the amplitude increases. This effect leads to a frequency instability area, which is further confirmed by performing reverse frequency sweeps. Figure~\ref{hyst}b compares upward and downward frequency sweeps for pillar $L1$, situated closer to the excitation source at 20~dBm. A typical hysteresis cycle is observed, showing an amplitude jump. This behavior confirms the presence of non-linearities in the proposed system, a rather unexpected result given the geometrical characteristics and the operating frequencies of the involved mechanical resonators. 

\subsection{Circularly-polarized resonances. }
\begin{figure}[t]
 \centering\includegraphics[width = 88mm]{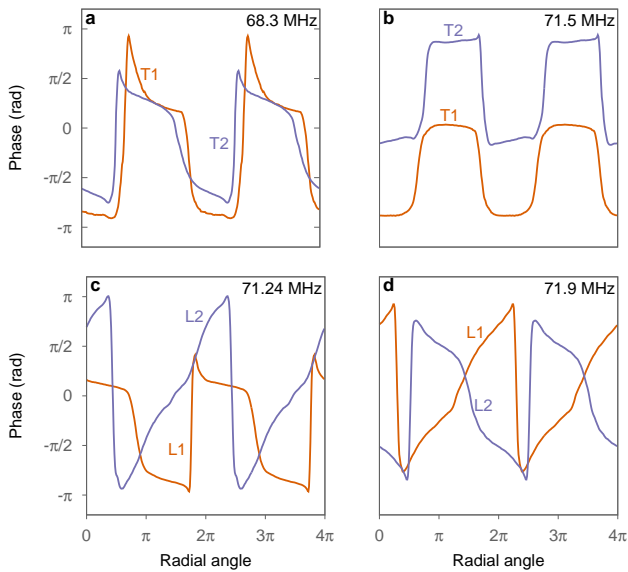}
 \caption{Typical phase states for the two proposed excitation schemes. The measurements are taken at a drive power of 20~dBm. (a) Symmetric mode at 68.3~MHz and (b) anti-symmetric mode at 71.5~MHz for a transverse excitation. (c) Illustration of rotating polarization states obtained for one out of two resonators in the case of a longitudinal excitation ($f=71.24$~MHz) and (d) for both resonators, with opposite handedness for an excitation frequency of 71.9~MHz. 
 } \label{RadPhase}
\end{figure}

The experimental field maps of Figure~\ref{15_long_nl}c further reveal that the input SAW amplitude also affects the nature of the resonator polarization states and breaks the flexural mode symmetry previously observed in transversely-coupled resonators. In the transverse excitation case reported in Figure~\ref{15_transv_nl}, only pure flexural modes are observed and the out-of-plane component of the displacement field exhibits a well-defined nodal line, as seen in the phase maps in Figure~\ref{15_transv_nl}c. The corresponding phase states are further depicted in Figure~\ref{RadPhase}a and b that represent the radial phase of the two resonators extracted from these field maps for an input power of 20~dBm. The two resonators are shown to oscillate either in-phase (Fig.~\ref{RadPhase}a) or out-of-phase (Fig.~\ref{RadPhase}b) depending on the considered vibration mode. But a longitudinal excitation leads to the occurrence of circular polarization states. In the case of mode~(\romannumeral 3), only one out of two resonators is circularly polarized, while the second one keeps a well-defined flexural behavior (Fig.~\ref{RadPhase}c). The phase state is table over the power range, as shown in the phase maps labeled (1) and (2) in Figure~\ref{15_long_nl}c taken at a drive power of 12~dBm. The higher-frequency mode~(\romannumeral 4), that appears at a frequency of about 72.9~MHz at 20~dBm, is however characterized by circularly-polarized states with opposite handedness (Figure~\ref{RadPhase}d). This behavior points at a cross-coupling between the two orthogonal polarization states theoretically composing the first-order flexural mode of such resonators with cylindrical symmetry. 


%
%
The occurrence of these circular states is concomitant with an increase in drive power; they may not, however, be indisputably be attributed to nonlinear effects. If non-linearities were already shown to induce elliptical polarization states in singly-clamped mechanical resonators~\cite{Nayfeh1995}, e.g. in carbon nanotubes~\cite{Perisanu_PRB2010,Vincent_JAP2019,Conley_NL2008}, the elliptical transition is obtained at very high drive powers and can only been obtained out of high quality-factor mechanical resonators. For low \textit{Q}-factors, this elliptical transition is considered trivial and hence independent of any non-linear behavior. But linear systems can also yield circular polarization states, provided that the resonators are subjected to force field with high vorticities~\cite{Gloppe_NNano2014}. In the present case, the nature of the Rayleigh SAW involves a linearly polarized excitation in the substrate plane. The vibration of the phononic resonator may however results in a mechanical back-action on the surface acoustic wave field at the resonator vicinity, resulting in coupling of the shear polarization components of the surface displacements. This, in turn, induces cross-coupling of the two orthogonal polarizations of the resonator, hence giving rise to the observed circular polarization states. This back-action process is, again, triggered by the higher force field induced by the cavity formed in the resonator-to-resonator gap, therefore leading to a power-dependant mechanism.

In conclusion, we demonstrated the occurrence of surface-acoustic wave induced non-linearities in pairs of otherwise linear coupled phononic microresonators. These non-linearities are characterized by a frequency softening of the pillar pair response for short gap distances, while isolated resonators retain a linear behavior. The characteristic features of the non-linear response are also conditioned by the incident SAW wave vector direction. Coupled resonators submitted to a transverse SAW excitation are well described by assuming a negative, cubic Duffing-like term applied to the resonator-to-resonator coupling constant, leading to a linear coupling rate of about 5~kN/m and a nonlinear coupling constant of the order of 10$^{-2}$~kN/m$^3$. In the case of a longitudinal excitation, in addition to the observed nonlinear dynamics, the increased surface displacements at the resonator vicinity results in a feedback mechanism between the mechanical motion of SAW-coupled phononic resonators and the substrate surface that acts as both a clamping element and a source of mechanical excitation. This interaction leads to a cross-coupling of the resonator eigenmodes, and hence on the occurrence of rotating polarization states. 
These results illustrate the rich dynamics involved in these phononic systems, where linear and non-linear effects can be combined to achieve coherent control of both the frequency response and modal behavior of high-frequency phononic microresonators.

\section*{Methods}
The phononic resonators under study consists of ion-beam-induced deposited platinum (IBID-Pt) cylinders. These pillars are deposited on a lithium niobate substrate (LiNbO$_3$ in the Y-crystallographic orientation. The substrate surface hosts a set of interdigitated transducers (IDTs) allowing for Rayleigh SAW excitation. The IDTs are made of a array of aluminum electrodes deposited following a chirped pattern, with a mechanical period varying from 34\mum{} to 68\mum, leading to the excitation of a Rayleigh wave over a frequency spectrum centered at about 73~MHz with a fractional bandwidth of 80\%. The IDTs are driven by a signal synthesizer allowing to control the radio-frequency (RF) signal power from -140 to +20~dBm, leading to the generation of a propagating Rayleigh wave with an out-of-plane displacement field reaching a maximum of 0.6~nm at 20~dBm at the substrate surface. 
These propagating Rayleigh waves are in turn used to excite the phononic resonators. The amplitude and phase of the out-of-plane component of the elastic displacement field is then imaged by optical interferometry using a Mach-Zehnder heterodyne laser scanning interferometer. 
The resonator geometry is therefore dictated by the trade-off imposed by the need to perform optical characterizations under a Rayleigh SAW excitation, imposing diameters in the micron-range and aspect ratio of the order of one, at least for the plain cylindrical geometry considered in the present work. In the present case, the resonator diameter was set to 4.4~$\mu$m, the height is comprised between 4.0 and 4.2\mum{} depending on the considered sample. As SAW-induced resonator coupling is expected to lead to altered coupling schemes as a function of both the separation distance and the chosen excitation conditions~\cite{Raguin_NComm2019}, the behavior of single resonators and of resonator pairs with gap distances $w$ equal to 1.5\mum{} 6\mum{} were investigated. Both transverse  and longitudinal excitation scheme were considered, where \textit{transverse} denotes an elastic wave front orthogonal to the inter-resonator axis while  \textit{longitudinal} refers to an elastic wave front parallel with the inter-resonator axis.

\section*{References}

\section*{Acknowledgements}
This project has received funding from the European Research Council (ERC) under the European Union's Horizon 2020 research and innovation programme (grant agreement No 865724). This work has also received support from the Agence Nationale de la Recherche under grant ANR-14-CE26-0003-01-PHOREST and the French RENATECH network with its FEMTO-ST technological facility. A. J. was supported by the EIPHI Graduate School (contract ``ANR-17-EURE-0002'').


%
%
%
%
\end{document}